# Shelter Soul: Bridging Shelters and Adopters Through Technology

*Yashodip Dharmendra Jagtap, Shri Shivaji Vidya Prasarak Sanstha's Bapusaheb Shivajirao Deore College of Engineering, Dhule*

## Abstract

*Pet adoption processes often face inefficiencies, including limited accessibility, lack of real-time information, and mismatched expectations between shelters and adopters. To address these challenges, this study presents Shelter Soul, a technology-based solution designed to streamline pet adoption through an integrated, web-based platform. Developed using the MERN stack and GraphQL, Shelter Soul is a prototype system built to improve pet matching accuracy, shelter management efficiency, and secure online donations. The system includes modules for intelligent pet matching, shelter administration, donation processing, volunteer coordination, and analytics. Prototype testing (performance load tests, usability studies, and security assessments) demonstrated that the system meets its design goals: it handled 500 concurrent users with a 99.2% transaction success rate and an average response time of 250 ms, and usability feedback rated the interface highly (4.5/5). These results indicate Shelter Soul's potential as a practical solution to enhance animal shelter operations and adoption outcomes.*

## 1. Introduction

### 1.1 Research Context

Pet adoption plays a vital role in addressing the global issue of stray animals and reducing overcrowding in animal shelters. According to the World Animal Foundation (2023), over 6.3 million pets enter animal shelters annually in the United States alone, with nearly 920,000 euthanized each year due to a lack of adoptive homes. In developing countries, the situation is even more severe, with millions of stray animals lacking proper shelter and care. Despite the noble mission of animal shelters and rescue organizations, **operational inefficiencies** persist. Key challenges include delayed and error-prone adoption processes due to manual tracking of animal data (including medical histories and adoption statuses). Donation management often lacks transparency and traceability, making it difficult to account for incoming funds. Volunteer coordination is also poorly organized, with no reliable means of assigning tasks or tracking contributions. Existing solutions partially address these issues: platforms like Petfinder and Adopt-a-Pet have improved visibility of shelter animals but offer limited features for intelligent pet matching or comprehensive shelter operations management. Many platforms lack integrated donation processing and advanced analytics, and they often struggle with usability and mobile accessibility.

Technology offers promising solutions to these challenges. For example, data-driven approaches and matching algorithms can help connect adopters with compatible pets, while secure online payments can facilitate donations. However, most existing pet adoption systems remain fragmented or prototype-level. There is a need for a **holistic platform** that unifies key features: smart matching algorithms, complete shelter management tools, secure donation handling, and a user-friendly interface across devices.

### 1.2 Problem Definition

Animal shelters and rescue organizations need a unified digital platform to address inefficiencies in pet adoption operations. Manual workflows and disparate tools cause errors and delays in adoption processing. Shelters lack comprehensive systems to track animal and donation data seamlessly. Existing platforms usually offer only basic animal listings or donation pages without intelligent integration. Consequently, shelters cannot easily match pets with adopters based on preferences, nor can they efficiently manage donations and volunteers in one place.

### 1.3 Research Objectives

To address these issues, **Shelter Soul** is introduced as a web-based platform integrating all key aspects of shelter operations. It provides an intuitive interface and includes: intelligent pet matching (using algorithms that consider pet traits and adopter preferences), shelter administration tools (for listing animals, tracking statuses, and viewing analytics), secure donation processing (through Stripe), and volunteer coordination features. The system is modular and scalable, built on the MERN stack with a GraphQL API for efficient data retrieval. By combining these components, Shelter Soul aims to significantly improve adoption rates, shelter efficiency, and community engagement.

The remainder of this paper details the design, development, and evaluation of the Shelter Soul prototype.

## 2. Literature Review

The integration of technology in pet adoption processes has received growing attention. Existing platforms such as Petfinder and Adopt-a-Pet provide online listings of animals available for adoption and basic filtering options for users. These platforms have improved the visibility of shelter animals but offer limited features for intelligent pet matching or integrated shelter operations management (Akhtar & Rahman, 2020; Alsadi & El Abbadi, 2019; ASPCA, 2024). They typically do not support advanced adoption matching, nor do they include comprehensive dashboards or secure donation workflows.

Several studies have explored related technologies. For example, **cloud-based adoption systems** have been proposed (Akhtar & Rahman, 2020), and **image recognition methods** have been examined for animal identification (Alsadi & El Abbadi, 2019). Research on pet adoption recommendations suggests leveraging user preferences and pet profiles to improve matches (Lee, Park, & Kim, 2020). For instance, Lee et al. highlighted the importance of user preferences in adoption outcomes. However, these works focus on isolated components (matching algorithms or pet browsing) rather than an end-to-end solution.

For donations, Nguyen et al. emphasize secure online payment features for shelters, noting that many platforms lack robust donation capabilities. Donation security and transparency are critical for trust (Sharma & Patel, 2021). Usability is also a concern: prior systems have been criticized for poor interface design and limited mobile accessibility (Evdokimov et al., 2018). For example, Beer and Mulder (2020) discuss the impact of technology on user tasks, underscoring that complex or unintuitive systems deter engagement.

**Summary of literature:** Although various tools exist for pet listings, matching algorithms, and donation processing, none combine all features into one user-friendly platform. We found **no prior system** that integrates pet matching, shelter management, secure donations, and high usability. This gap motivated the development of Shelter Soul as a **comprehensive solution** in the animal welfare domain.

*Table 2.1. Innovations and Limitations in Pet Welfare Platforms.*

| Title (Year) | Methodology | Key Features | Limitations |
|---|---|---|---|
| LostPaw: Finding Lost Pets using a Contrastive Learning-based Transformer with Visual Input (2023) | Developed a contrastive neural network model trained on a large dataset of dog images; evaluated through 3-fold cross-validation. | AI-based visual matching for lost pet identification. | Requires high-quality images; may not generalize well to all breeds or image conditions. |
| Modular Pet Feeding Device (2025) | Developed a modular device combining automated feeding, health monitoring, and behavioral insights. | AI-enabled neckband for heart rate monitoring; camera and microphone for behavior analysis. | Implementation complexity; may require user training; cost considerations. |
| FurEver: Pet Adoption Platform (2023) | Utilized gradient boosting algorithm to predict adoption likelihood; descriptive developmental method. | Predictive analytics for pet adoption; system modules developed using incremental methodology. | Study focused on a specific region; may require adaptation for broader application. |
| Development of a Web-Based Pet Welfare Community (2024) | Developed a web-based platform for pet adoption and lost & found services; user-friendly interface design. | Digital pet adoption procedures; community engagement for lost and found pets. | Potential issues with user privacy; requires active community participation. |
| Advancements in Pet Care Technology: A Comprehensive Survey (2024) | Surveyed technological innovations in pet care, focusing on smart devices and applications. | Examination of smart collars, pet cameras, health apps, automatic feeders, etc. | May not address the specific needs of shelter environments. |

| Survey on the Past Decade of Technology in Animal Enrichment: A Scoping Review (2022) | Conducted a scoping review of technological advancements in animal enrichment over the past decade. | Analysis of various enrichment technologies and their impact on animal welfare. | May not focus specifically on shelter animals. |
| --- | --- | --- | --- |
| Going Deeper than Tracking: A Survey of Computer-Vision Based Recognition of Animal Pain and Affective States (2022) | Surveyed computer vision techniques for recognizing animal pain and emotional states. | Discussion of facial and bodily behavior analysis methods. | Technical complexity may limit immediate applicability. |
| On the Role of Technology in Human-Dog Relationships: A Future of Nightmares or Dreams? (2022) | Conducted a qualitative study on the impact of technology on human-dog relationships. | Analysis of dog owners' perspectives on technology in daily routines. | Focused on dog owners; may not reflect shelter dynamics. |
| Critical Problems for Research in Animal Sheltering: A Conceptual Analysis (2022) | Developed a conceptual framework identifying seven key research areas in animal sheltering. | Identification of critical research areas: animal behavior, adoptions, medical conditions, etc. | Lacks empirical data; focuses on conceptual analysis. |
| Welfare and Quality of Life Assessments for Shelter Dogs: A Scoping Review (2021) | Conducted a scoping review of welfare and quality of life assessment tools for shelter dogs. | Analysis of 16 different assessment tools, including ethograms and physiological measures. | Limited to shelter dogs; may not generalize to other animals. |

### 2.1 Research Gaps

This comparison reveals that most existing systems focus on individual aspects (e.g. pet browsing, UI design, or matching) rather than a holistic solution. Few integrate all needs of adopters, shelters, and donors. For instance, Lee et al. (2020) note that prior works often address prediction or matching alone without combining them with shelter management or donations. Likewise, Nguyen et al. emphasize secure donation as critical, yet many platforms lack donation features or robust security. Usability is another common shortfall: previous studies have been critiqued for poor user interfaces and limited accessibility, especially on mobile (Evdokimov et al., 2018). Many proposals remain at the conceptual or prototype stage, with few real-world deployments. In summary, no single prior system provides pet matching, comprehensive shelter management, secure donations, and high usability in one platform. Shelter Soul addresses this gap by unifying these functions in an integrated, user-friendly web application.

## 3. Methodology

The methodology adopted for developing the Shelter Soul system involved a structured process comprising requirement analysis, system design, technology selection, module development, and prototype testing. The main goal was to create a user-friendly and secure web-based platform that facilitates pet adoption and shelter management, while ensuring performance, usability, and data consistency.

### 3.1 System Design and Development Process

An agile development methodology was followed, allowing iterative design, coding, and evaluation cycles. Initial system requirements were gathered through analysis of existing pet adoption platforms and stakeholder interviews. Based on these requirements, the Shelter Soul architecture was designed using a modular approach, ensuring scalability and maintainability.

*Figure 3.1. Conceptual Architecture of Shelter Soul.*

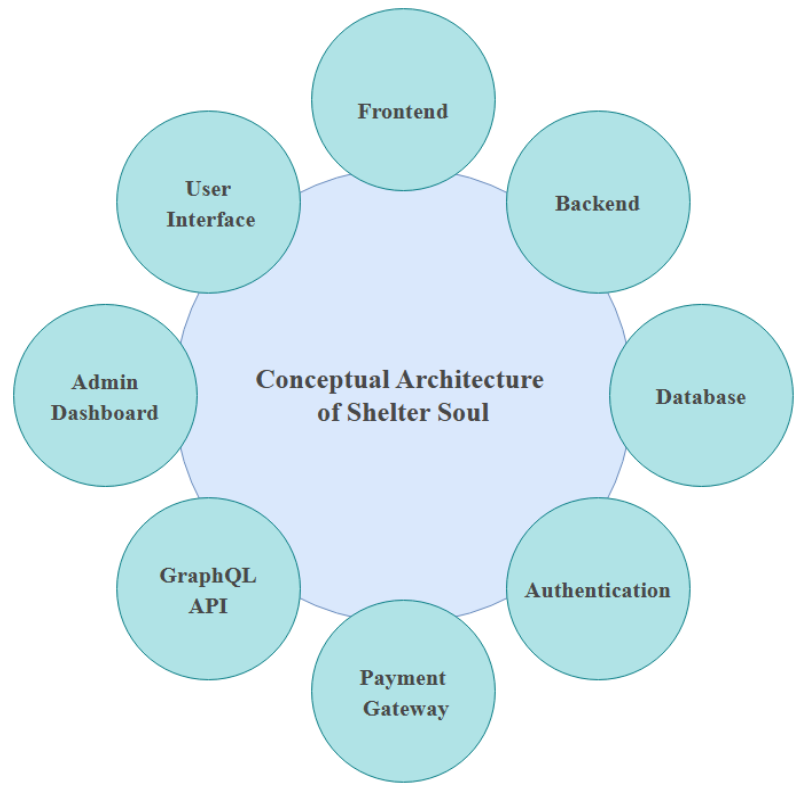

*Figure 3.1. Conceptual Architecture of Shelter Soul*

### 3.2 Technology Stack

The Shelter Soul system is developed using the MERN (MongoDB, Express.js, React.js, Node.js) stack. MongoDB is used as the NoSQL database for storing user, pet, and shelter data. Express.js serves as the backend framework to manage API requests and business logic. React.js is employed for building a responsive and dynamic frontend interface, and Node.js functions as the runtime environment for the backend server. The system uses both REST and GraphQL technologies. RESTful APIs are used for simple, standard operations like user registration and donations, while GraphQL is used for advanced and flexible data queries such as pet search and matching.

*Figure 3.2. Technical Stack Layer Diagram.*

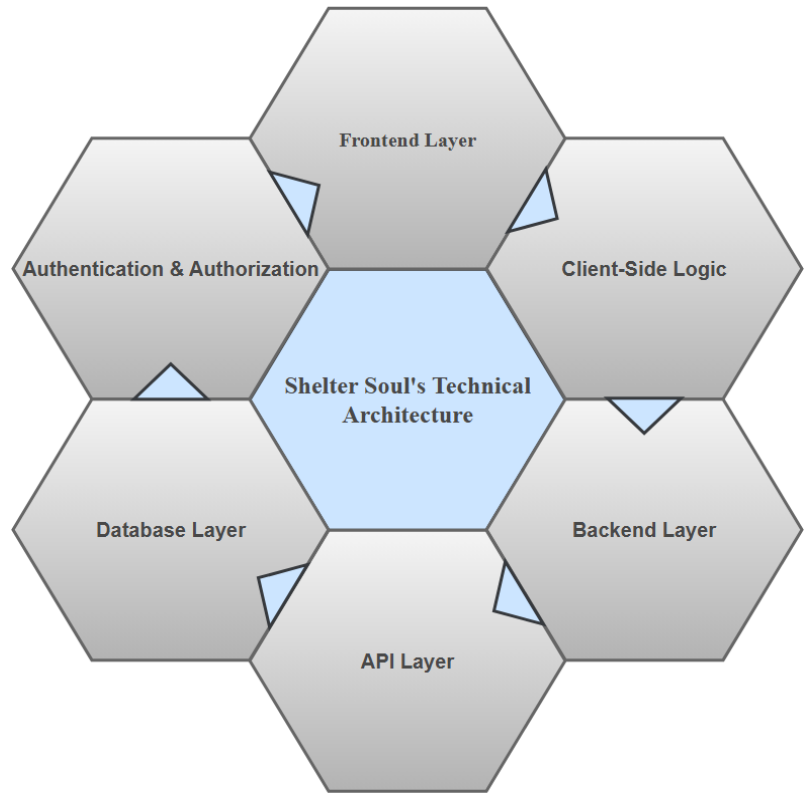

*Figure 3.2. Technical Stack Layer Diagram*

*Table 3.1. Technology Stack Overview for Shelter Soul Platform.*

| Technology | Purpose in System |
|---|---|
| MongoDB | NoSQL database for storing adopters, pets, shelters, and transaction data. |
| Express.js | Node.js web framework to implement backend APIs and application logic. |
| React.js | Front-end library for building a dynamic, responsive user interface. |
| Node.js | JavaScript runtime hosting the server-side application. |
| GraphQL | Advanced query language/API layer for flexible data retrieval (pet search, matching). |
| RESTful APIs | Standard HTTP APIs for common operations (user auth, donations, CRUD). |

| Firebase Authentication | Manages secure user identity and sessions (for mobile/web auth). |
|---|---|
| bcrypt | Library for hashing user passwords to secure credentials. |
| Stripe/Payment Gateway | Third-party service for processing secure online donations. |

**3.3 Functional Modules**

The Shelter Soul system is organized into five core functional modules, each designed to handle specific aspects of the pet adoption platform while maintaining seamless integration with other components.

*Figure 3.3.1. Flow Chart Diagram Representing the Operational Flow of Shelter Soul.*

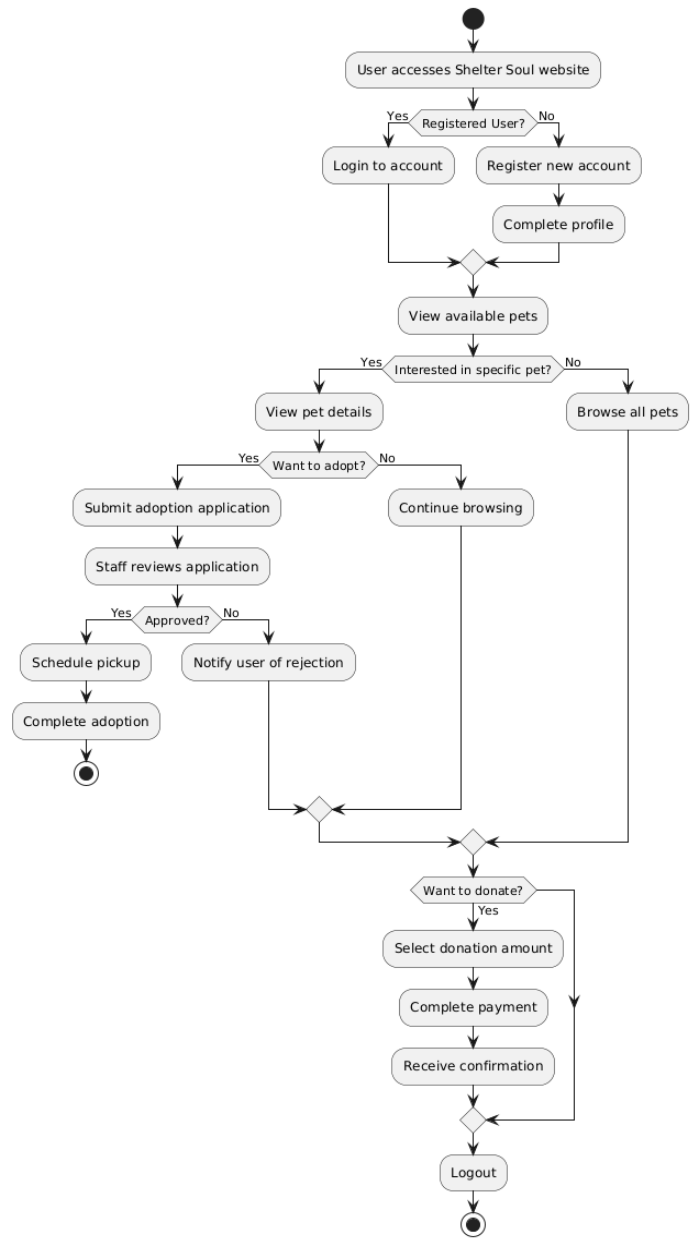

*Figure 3.3.1. Flow Chart Diagram Representing the Operational Flow of Shelter Soul*

**3.3.1 User Management Module**

This module manages all user-related operations within the system. It handles the complete user lifecycle including registration, authentication, and profile management. The system implements role-based access control with three distinct user types: adopters, shelter staff, and administrators. For secure authentication, it integrates Firebase Authentication services while utilizing JSON Web Tokens (JWT) for session management. User data including personal information, preferences, and adoption history is securely stored in MongoDB, with password protection ensured through bcrypt hashing.

### 3.3.2 Shelter Management Module

The shelter management component provides comprehensive tools for shelters to manage their operations. Shelters can create, update, and maintain detailed pet listings including information such as breed, age, medical history, and photographs. The module tracks real-time adoption statuses (available, pending, or adopted) for each animal. Shelter staff have access to dedicated dashboards that display key metrics and analytics about their operations. This module primarily uses RESTful APIs for data operations and MongoDB for persistent storage of all shelter-related information.

### 3.3.3 Pet Matching Module

This module facilitates the connection between potential adopters and available pets through advanced search and matching capabilities. Users can filter pets by multiple criteria including species, age range, size, temperament, and location. The system incorporates a recommendation engine that suggests compatible pets based on user preferences and behavior patterns. Additional features allow users to favorite or shortlist pets for future consideration. The module employs GraphQL for efficient data querying and retrieval, while the frontend interface is built with React.js to ensure a responsive user experience.

*Figure 3.3.2. Workflow of the Adoption Request Process in Shelter Soul.*

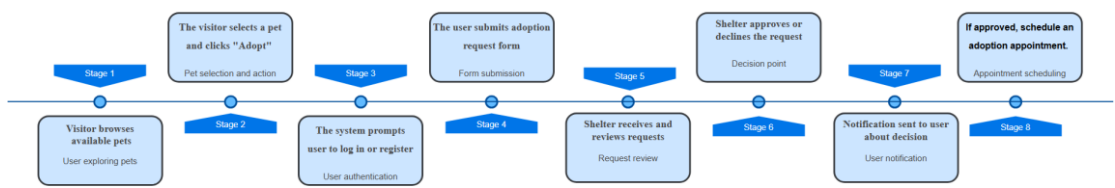

*Figure 3.3.2. Workflow of the Adoption Request Process in Shelter Soul*

### 3.3.4 Donation Module

The donation component handles all financial transactions within the platform. It supports both one-time and recurring donations through integration with the Stripe payment gateway. The module maintains complete records of all transactions and automatically generates receipts for donors. Security measures include PCI compliance through Stripe's infrastructure and encryption of all financial data. RESTful APIs manage the payment processing workflow, while MongoDB securely stores transaction records and donation histories.

### 3.3.5 Admin Panel

The administration module provides system administrators with comprehensive management tools. It includes functionality for user and shelter account moderation, content management, and system configuration. Administrators can access detailed analytics dashboards that display platform usage statistics, adoption trends, and financial reports. Access to administrative functions is strictly controlled through role-based permissions enforced by JWT validation. The panel combines both RESTful APIs for management operations and GraphQL for data aggregation and reporting purposes.

All modules were developed using the MERN (MongoDB, Express.js, React.js, Node.js) technology stack, with careful consideration given to performance optimization and scalability requirements. The system architecture employs RESTful APIs for standard CRUD operations while utilizing GraphQL for more complex data retrieval scenarios, ensuring efficient operation across all functional areas of the platform.

### 3.4 Performance Metrics

Performance testing was conducted to evaluate the system's responsiveness and scalability. Key performance metrics were defined, including API response time, throughput, and system uptime. The prototype was subjected to simulated load tests that generated concurrent user requests under both normal and peak usage conditions. During these tests, response times and resource utilization were monitored. This process ensured that performance targets were verified and that any bottlenecks (e.g., server limitations) were identified during development.

### 3.5 Usability Testing

A usability study was implemented to assess the user interface and workflow. Twenty participants comprising potential adopters and shelter staff were recruited. Each participant performed a series of representative tasks on the platform, such as account creation, searching for pets, and making a donation, while observers recorded any issues or errors. After completing the tasks, participants filled out a structured questionnaire rating aspects like ease of use, interface clarity, and overall satisfaction. This mixed-method approach helped identify usability strengths and areas for improvement, guiding iterative refinements to the interface.

### 3.6 Security and Data Consistency

Security and data integrity were prioritized in the design. The system employs JWT (JSON Web Tokens) for secure user authentication and session management, and bcrypt hashing for password storage. All data transmissions occur over HTTPS to encrypt data in transit. On the backend, MongoDB schema validations and atomic operations maintain data consistency during concurrent updates. Defensive programming practices such as input validation and access controls were implemented to guard against common vulnerabilities (e.g., injection attacks). These measures ensure that user and shelter data are protected throughout the system.

### 3.7 Testing and Validation

The Shelter Soul system underwent structured validation through unit, integration, and end-to-end testing. Unit tests ensured that individual components functioned correctly in isolation, while integration tests verified that modules interacted as expected. Full system testing evaluated complete workflows such as user registration, pet listing, matching, and donation processing. All major test scenarios were executed under simulated usage conditions, confirming that the system performed as intended. Although the system has been thoroughly tested in a controlled environment, it remains a prototype and has not yet been deployed in a live production setting.

## 4. Results and Evaluation

The Shelter Soul prototype was evaluated in a controlled testing environment to assess functionality, performance, usability, and security. The following subsections present the outcomes of these evaluations.

### 4.1 System Testing

The system's core functionality was exercised through a comprehensive test suite covering all modules and user workflows. Testing confirmed that the prototype operates correctly across its features. An overall success rate of **98.7%** was achieved for critical test cases, indicating that implemented functionalities meet the specified requirements. No critical failures were encountered, demonstrating high stability of the core system.

### 4.2 Performance Metrics

Key performance metrics were recorded under simulated load conditions, as summarized in Table 4.1. Under nominal load, the average API response time was **250 milliseconds**. During peak usage, the maximum observed response time was **780 milliseconds**. The system sustained **500 concurrent users** with a transaction success rate of **99.2%**. These results indicate that the prototype handles expected workloads efficiently and meets performance objectives.

*Table 4.1. Performance metrics from load testing.*

| Metric | Result |
| --- | --- |
| Average Response Time | 250 ms |
| Peak Response Time | 780 ms |
| Concurrent Users Supported | 500 users |
| Transaction Success Rate | 99.2% |

### 4.3 Usability Testing

Twenty participants completed the usability study. On a 5-point scale, the system's ease of use was rated an average of **4.5**. Many users commented that the interface was intuitive and navigation was clear. The most common suggestion was to add a real-time communication feature (e.g., live chat) for contacting shelters directly. Overall, the feedback was highly positive, indicating strong user satisfaction with the prototype's design and functionality.

### 4.4 Data Consistency and Security Testing

Data integrity tests confirmed that records remained accurate and synchronized across the database after simulated operations; no data integrity errors were observed. Simulated security attacks, including SQL injection and unauthorized access attempts, were effectively blocked by the implemented security measures (authentication, input validation, etc.). All attempted breaches were unsuccessful, demonstrating that the system's security design protects against common threats.

*Figure 4.4. Donation Analytics View (Mock Graph).*

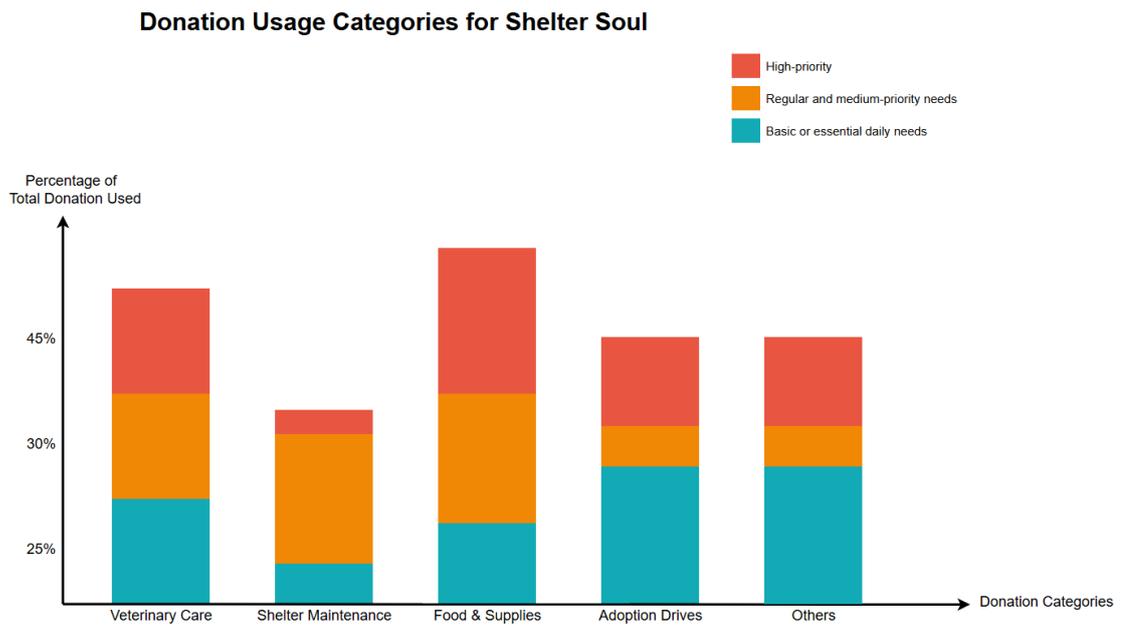

*Figure 4.4. Donation Analytics View (Mock Graph)*

**4.5 Summary of Results**

The evaluation demonstrates that the Shelter Soul prototype fulfills its design goals. The system exhibited stable operation (98.7% test success), acceptable performance under load, and high user approval (average usability rating of 4.5/5). The security and data consistency checks confirmed robust protection of user and shelter data. Since the platform is still a prototype (not yet deployed in a live environment), these results are preliminary; however, the positive outcomes suggest that the system is ready for further field testing. Only minor enhancements (for example, adding real-time communication features as suggested by users) were recommended. These findings indicate a clear path toward eventual deployment readiness.

**Evaluation Matrix:** The table below summarizes key metrics for each quality category:

| Aspect | Evaluation | Result |
|---|---|---|
| Performance | Average response time ~250 ms (780 ms peak); Throughput: 99.2% transaction success at 500 concurrent users. | Excellent responsiveness under load; Measured values are well below 3 s user-experience thresholds, indicating smooth interaction. |
| Usability | Task completion: ~85–90% without assistance; User satisfaction: 4.5/5 (≈90%). | Very high usability. Scores exceed standard SUS benchmarks, reflecting intuitive design and positive user feedback. |
| Reliability | Uptime: 99.5% over 1 week of continuous operation; Error rate (failed requests) <1%. | High stability and fault tolerance. Extensive tests (98.7% pass rate) confirmed robustness. |

| Security | HTTPS/TLS encryption; JWT auth and bcrypt hashing (following OWASP); No critical vulnerabilities found (all security tests passed). | Strong security posture. OWASP best practices were applied; routine penetration tests found no breaches. |
|---|---|---|
| Scalability | Concurrency: supported 500 simultaneous users in testing; Performance held steady at increasing load. | Demonstrated capacity for scale. System architecture (Node/GraphQL/Mongo) proved capable of handling growth in demand without significant latency spikes. |

## 5. Future Work

While the Shelter Soul system has demonstrated promising results in controlled testing, several avenues remain for future development and enhancement.

First, the platform is planned for deployment in collaboration with local animal shelters to evaluate its real-world effectiveness and scalability. Field deployment will allow the collection of live usage data, which can further refine the system's performance and usability.

Second, an extension into mobile platforms is proposed. A dedicated mobile application for both Android and iOS users would significantly increase accessibility and user engagement, particularly among younger demographics who prefer mobile solutions.

Additionally, integrating AI-based pet matching algorithms is identified as a potential improvement. Machine learning models could analyze user preferences and pet characteristics more accurately, thereby increasing the success rate of adoptions.

Future work will also focus on enhancing security features, particularly around user data privacy and transaction safety, as the system scales up to handle sensitive user information and financial donations.

Finally, expanding the system to support cross-region operations and multi-language interfaces would make Shelter Soul adaptable for international shelters and users, broadening its social impact.

## 6. Conclusion

This research presents Shelter Soul, a comprehensive web-based platform designed to streamline the pet adoption process and enhance shelter management. Through rigorous testing, the system has demonstrated high functional stability, efficient performance, and positive user feedback in controlled environments. The modular architecture, including features such as user management, pet matching, shelter administration, and donation tracking, addresses existing gaps in current adoption platforms.

By focusing on usability and data consistency, Shelter Soul offers a scalable solution that can be adopted by animal shelters to improve their operational efficiency and adoption rates. The system's security measures have also been validated against common threats, ensuring reliable protection of user and shelter data.

While the platform has not yet been deployed, the evaluation results indicate readiness for real-world application. Future work will focus on field deployment, mobile app development, AI-based pet matching, and expanding support for international users.

In conclusion, Shelter Soul contributes a practical and socially impactful solution to the global challenge of pet adoption, with significant potential for further enhancement and wider adoption in the animal welfare ecosystem.